\documentclass[twocolumn]{jpsj2} 
\title{Isotropic, Nematic and Smectic A Phase Behaviour in a Fictitious Field} 
\author{
Masashi  \textsc{Torikai}\thanks{E-mail address: torikai@phen.mie-u.ac.jp}
and
Mamoru \textsc{Yamashita}\thanks{E-mail address: mayam@phen.mie-u.ac.jp}}

\inst{Department of Physics Engineering, Faculty of Engineering, Mie
University, Kamihama 1515, Tsu 514-8507}

\abst{
Phase behaviours of liquid crystals under external fields, conjugate to
the nematic order and smectic order, are studied within the framework of
mean field approximation developed by McMillan.
It is found that phase diagrams, of temperature vs interaction parameter
of smectic A order, show several topologically different types caused by
the external fields.
The influences of the field conjugate to the smectic A phase, which is
fictitious field, are precisely discussed.
}

\kword{nematic phase, smectic A phase, McMillan model, phase transition,
external field}

\begin{document}
\maketitle

\section{Introduction}
Electric and magnetic fields may influence on the liquid crystalline
ordering and thus on the phase behaviour of the liquid crystal systems.
These fields directly couple to the nematic order and influence on the
nematic ordering. 
Strong external fields may induce a nematic order even
in the isotropic phase~\cite{wojtowicz,Lelidis1993b}.
The external fields also slightly raise the isotropic-nematic (I-N)
transition temperature.
As the external field is increased, the nematic order in the nematic
phase at the transition temperature decreases; 
together with the increase of the nematic order in the isotropic phase
at the transition temperature, the discontinuity at the I-N transition
is reduced.
Eventually the discrete I-N transition vanishes under a sufficiently
strong field, as shown in ref.~\citen{wojtowicz} theoretically and in
ref.~\citen{Lelidis1993b} experimentally. 

The influences of the external fields show a richer variety and are much
complex if we consider the smectic A order in addition to the nematic 
order.
For example, a system which intrinsically shows direct isotropic-smectic
A (I-A) transition under zero external field can exhibit the
nematic phase between isotropic and smectic A phases under an external
field. 
The existence of such a field-induced (non-spontaneous) nematic phase
was predicted theoretically in ref.~\citen{Rosenblatt1981a} and
found experimentally in ref.~\citen{Lelidis1994}.
The appearance of this non-spontaneous nematic phase is a consequence of
the direct coupling between the field and the nematic order.
The external fields also couple to the smectic A order indirectly,
through the direct coupling between nematic order and smectic A order.
Thus the fields may influence the smectic order and may change the type
of the nematic-smectic A (N-A) transition.
Actually, it is shown that a system which shows a first order N-A
transition under zero external fields may undergo a second order N-A
transition under an external field~\cite{Rosenblatt1981b,Hama}.

These studies have investigated the effects of the external field
conjugate to the nematic order, since the (squares of) magnetic and
electric fields are the fields conjugate to the nematic
order~\cite{deGennes}. 
The external field conjugate to the smectic order, which should induce a
one-dimensional periodic density modulation in addition to the
orientational order of the molecules, is not experimentally available
and not realistic field.  
(The electric field may play a role of the field conjugate to
the \emph{smectic order} by the effect of, so called, smectic
electrostriction~\cite{Lelidis1996}, which is the density change induced
by the electric field. 
However, the effect of the electrostriction is negligibly weak 
except for the case where the coupling between nematic order and
smectic order is weak~\cite{Lelidis1996}.)

The purpose of the present paper is to clarify the phase behaviour of
the liquid crystals under the influence of not only the external field,
$h_{s}$, conjugate to the nematic order but also the external field,
$h_{\sigma}$, conjugate to the smectic order.
The field $h_{\sigma}$ is fictitious and arbitrary field;
however, the application of such an external field is of interest from
theoretical aspects since it completes the description of the phase 
diagram. 
In addition, a field conjugate to an order parameter naturally arises as
an effective field, as a result of the spatial inhomogeneity of the
order parameter. 
Such an effective field is helpful to understand the phase transitions
in inhomogeneous systems.
In fact, one of the authors has disclosed the mechanism of phase
transition of the freely suspended smectic films, by utilising  the
effective field conjugate to the smectic order
parameter~\cite{yamashita2003a}.

In the present paper, we investigate the McMillan's liquid crystal 
model~\cite{McMillan}, which has a parameter $\alpha$ showing a
dimensionless interaction strength for the smectic A phase.
By analysing order parameters and transition temperatures for a
suitable range of $\alpha$, we elucidate the phase behaviour of the
system.
The knowledge of the behaviour of the homogeneous systems under the
fictitious field is useful to investigate the inhomogeneous systems;
the relation between the fictitious field and the effective field due to
the spatial inhomogeneity is discussed in \S4.

\section{McMillan Model}
We use a mean field approximation of liquid crystalline model introduced
by McMillan~\cite{McMillan}.
Let $s$ and $\sigma$ be order parameters of nematic order and smectic
order, respectively.
Then, the McMillan's one-body potential is expressed as
\begin{equation}
 V(z, \cos \theta)=
  -V_{0}
  \left[
   s+\alpha \sigma \cos\left(\frac{2\pi z}{d}\right)
  \right]
  P_{2}(\cos \theta), \label{eq:McMillanPotential}
\end{equation}
where we assume the director is parallel to $z$-axis, and $\theta$ is
the angle between the long axis of a molecule and $z$-axis;
$P_{2}(x)=3(x^{2}-1)/2$ is the second order Legendre polynomial 
and $d$ denotes the thickness of a single smectic layer.
The second term on right hand side of eq.~\eqref{eq:McMillanPotential}
denotes the smectic order interaction and parameter $\alpha$ is the
smectic order interaction strength. 
The order parameters are self-consistently determined from
\begin{align}
 s &=\langle P_{2}(\cos \theta) \rangle_{\beta, s, \sigma}, \\
 \sigma &=
 \left\langle
 \cos\left(\frac{2\pi z}{d}\right)P_{2}(\cos \theta)
 \right\rangle_{\beta, s, \sigma},
\end{align}
where the angular brackets denote the canonical ensemble average for
one-particle: 
\begin{multline}
 \langle A(\theta, z) \rangle_{\beta, s, \sigma} \\
 \begin{split}
  = & \frac{1}{Z(\beta, \beta V_{0} s, \beta V_{0} \alpha \sigma)}
  \int_{0}^{\pi} d\theta \sin \theta \int_{0}^{d} dz 
  A(\theta, z) \\
  & \times \exp\left\{
  \beta V_{0} \left[
  s+\alpha \sigma \cos\left(\frac{2\pi z}{d}\right)
  \right]
  P_{2}(\cos \theta)
  \right\}, 
 \end{split}
\end{multline}
where $Z$ denotes the one-particle partition function
\begin{multline}
  Z(\beta, \beta V_{0}s, \beta V_{0}\sigma) \\
 \begin{split}
  = & \int_{0}^{\pi} d\theta \sin \theta \int_{0}^{d} dz \\
  & \times \exp\left\{
  \beta V_{0} \left[
  s+\alpha \sigma \cos\left(\frac{2\pi z}{d}\right)
  \right]
  P_{2}(\cos \theta)
  \right\}. \label{eq:partitionFunction}
 \end{split}
\end{multline}
Under the external fields, the potential energy induced by the external
fields is
\begin{equation}
 V_{\text{ext}}=
  -\left[
   h_{s} P_{2}(\cos \theta)
   + h_{\sigma}
   \cos \left(\frac{2\pi z}{d}\right) P_{2}(\cos \theta)
  \right],
\end{equation}
where $h_{s}$ and $h_{\sigma}$ are the external fields conjugate to,
respectively, $s$ and $\sigma$.
The mean field one-particle partition function is then
$Z(\beta,\beta h_{s}+\beta V_{0} s, \beta h_{\sigma}+\beta V_{0} \alpha 
\sigma)$.
For later convenience to obtain the free energy, we introduce symmetry
breaking fields~\cite{Nakano} $\eta$ and $\zeta$, 
which are conjugate to the nematic order and smectic order,
respectively. 
The symmetry breaking fields should be zero at the thermal equilibrium.
The one-particle partition function with symmetry breaking fields is
$Z(\beta,\eta+\beta h_{s}+\beta V_{0} s, \zeta+\beta h_{\sigma}+\beta
V_{0} \alpha \sigma)$. 
The self-consistent equations for order parameters are
\begin{subequations}
 \begin{align}
  s=
  I(\beta,
  \eta+\beta h_{s}+\beta V_{0} s,
  \zeta+\beta h_{\sigma}+\beta V_{0} \alpha \sigma),
  \label{eq:sWithField}\\
  \sigma=
  J(\beta,
  \eta+\beta h_{s}+\beta V_{0} s,
  \zeta+\beta h_{\sigma}+\beta V_{0} \alpha \sigma),
  \label{eq:sigmaWithField}
 \end{align} \label{eq:ssEqWithField}
\end{subequations}
where $I(\beta, \eta, \zeta)$ and $J(\beta, \eta, \zeta)$ are defined as
\begin{subequations}
 \begin{align}
  I(\beta, \eta, \zeta)=\frac{\partial}{\partial \eta}
  \ln Z(\beta, \eta, \zeta),
  \\
  J(\beta, \eta, \zeta)=\frac{\partial}{\partial \zeta}
  \ln Z(\beta,\eta,\zeta).
 \end{align}\label{eq:IandJ}
\end{subequations}
Among the sets of solutions of eqs.\eqref{eq:ssEqWithField}, the
thermodynamically stable set gives a minimum of the following function 
(the work done by the symmetry breaking fields):
\begin{multline}
 \beta \tilde{F}(\beta, h_{s}, h_{\sigma}; s,\sigma)\\
  =\int_{0}^{s}ds' \eta(h_{s}, h_{\sigma}; s',0)
  +\int_{0}^{\sigma}d\sigma' \zeta(h_{s}, h_{\sigma}; s,\sigma'),
  \label{eq:pseudoFreeEnergy}
\end{multline}
where $\eta(h_{s}, h_{\sigma}; s, \sigma)$ and
$\zeta(h_{s}, h_{\sigma};s, \sigma)$ are the inverse solutions 
$\eta$ and $\zeta$ of the self-consistent equations
\eqref{eq:ssEqWithField}.
The minimum of $\tilde{F}(\beta, h_{s}, h_{\sigma}; s,\sigma)$ for given
$h_{s}$ and $h_{\sigma}$ is the thermodynamic free energy per molecule.
We can rewrite eq.\eqref{eq:pseudoFreeEnergy} by use of
eqs.\eqref{eq:ssEqWithField} as
\begin{multline}
 \beta
 \left(
 \tilde{F}(\beta, h_{s}, h_{\sigma};s,\sigma)
 -\tilde{F}(\beta, h_{s}, h_{\sigma};0,0)
 \right)\\
 \begin{split}
  =& \frac{\beta V_{0}}{2} s^{2}
  +\frac{\beta V_{0}}{2} \alpha \sigma^{2}\\
  &-\ln \frac{Z(\beta,\beta h_{s}+\beta V_{0} s,
  \beta h_{\sigma}+\beta V_{0} \alpha \sigma)}
  {Z(\beta, 0, 0)},
 \end{split}
\end{multline}
where $s$ and $\sigma$ are the solutions of eqs.\eqref{eq:ssEqWithField} 
for given $h_{s}$ and $h_{\sigma}$ with $\eta=\zeta=0$. 

\section{Results}
\begin{figure}
\begin{center}
 \includegraphics[width=8.5cm]{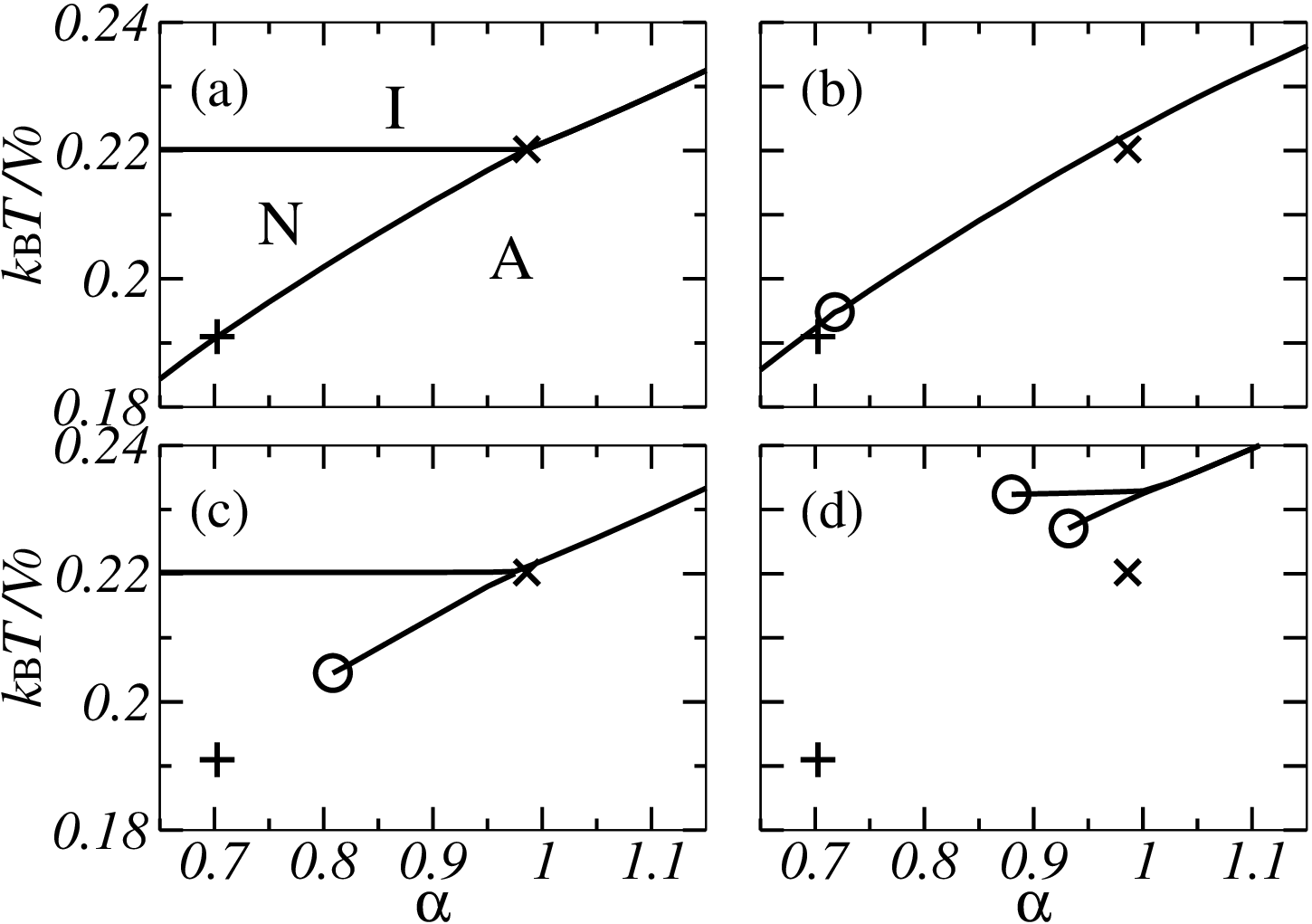}
\end{center}
\caption{
 Phase diagrams for
 (a)$h_{s}=0$, $h_{\sigma}=0$,
 (b)$h_{s}=0.0106>h^{\text(c)}$, $h_{\sigma}=0$,
 (c)$h_{s}=0$, $h_{\sigma}=0.003$,
 and (d)$h_{s}=0.0106$, $h_{\sigma}=0.025$.
 In (a), the isotropic, nematic and smectic A phases are indicated as
 I, N and A, respectively.
 The triple point and tricritical point are shown in (a) as $\times$ and
 $+$, respectively; 
 these points for (a) are also shown on (b), (c) and (d) for reference. 
 The circles indicate the tricritical point in (b), and the critical
 point in (c) and (d).
 } 
\label{fig:4phaseDiagrams}
\end{figure}

Here, we show our numerical results obtained from the
above formulation.
For convenience of the following observation, we first show the phase
diagram of the systems without external fields, which is originally
shown in ref.~\citen{McMillan}. 
The $\alpha$-$T$ phase diagram is shown in
Fig.\ref{fig:4phaseDiagrams}(a). 
For a system with small $\alpha$, i.e., $\alpha<0.986$, the system
undergoes I-N transition at $T=T_{\text{IN}}=0.22019$, and
N-A transition at lower temperature $T_{\text{NA}}$. 
The transition temperature $T_{\text{NA}}$ is a monotonically increasing
function of $\alpha$. 
For small $\alpha$, the N-A transition is a second order transition,
i.e., $s$ and $\sigma$ are continuous but their differential
coefficients have discontinuity at the N-A transition. 
If $\alpha$ and $T_{\text{NA}}$ are sufficiently large, i.e.,
$\alpha>\alpha_{\text{tc}}$ and $T_{\text{NA}}>T_{\text{tc}}$, the N-A
transition turns to first order transition.
The tricritical values of $\alpha$ and $T_{\text{NA}}$ are
$\alpha_{\text{tc}}=0.703$ and $T_{\text{tc}}=0.1910$.
The existence of the tricritical point on N-A coexisting line was also
pointed out within the framework of the Landau-de Gennes
theory~\cite{deGennes}. 
The N-A coexisting line meets I-N coexisting line at the triple point
$\alpha_{\text{triple}}=0.986$, $T_{\text{triple}}=0.2202$.
A system with $\alpha>\alpha_{\text{triple}}$ exhibits direct I-A 
transition.

We directly show that the independence of $T_{\text{IN}}$ on
$\alpha$ using the analogue of the Clapeyron relation for coexisting
lines of fluid phases~\cite{Callen}.
As done in the Clapeyron relation we can write the slope of a coexisting
line at a point $(\alpha, T_{\text{IN}}(\alpha))$ as
\begin{equation}
 \frac{d T_{\text{IN}}(\alpha)}{d \alpha}
  =
  -\frac{\Delta F_{\alpha}}{\Delta F_{T}},\label{Clapeyron}
\end{equation}
where $F$ is the thermodynamic free energy per molecule given by
minimizing $\tilde{F}(\beta, h_{s}, h_{\sigma}; s, \sigma)$ with respect
to $s$ and $\sigma$.
The factors on the right hand side of eq.\eqref{Clapeyron} are
determined as $\Delta F_{\alpha}=\partial F/\partial
\alpha|_{\text{nematic}} -\partial F/\partial\alpha|_{\text{isotropic}}$
and  $\Delta F_{T}=\partial F/\partial T|_{\text{nematic}}
-\partial F/\partial T|_{\text{isotropic}}$, where
$\partial F/\partial \alpha$ and $\partial F/\partial T$ are evaluated
at the point $(\alpha, T_{\text{IN}}(\alpha))$ and at the phase
indicated by the subscript.
Then $\Delta F_{\alpha}=-\Delta(V_{0}\sigma^{2}/2)=0$ since $\sigma$ is
zero in both isotropic and nematic phases;
$\Delta F_{T}$ is the entropy change at the transition and is nonzero.
Therefore $T_{\text{IN}}$ is independent of $\alpha$.

In the following, we investigate the phase diagrams under the influence
of external fields, $h_{s}$ and $h_{\sigma}$.
We restrict our investigations to ranges $0 \leq h_{s} \leq 0.03$
and $0 \leq h_{\sigma} \leq 0.06$.
We believe that the phase diagrams are qualitatively the same if the
fields exceed the ranges, and that these upper bounds are sufficiently
large to observe the influence of the fields. 
We also confirmed that, using our inhomogeneous model systems, the
effective fields induced by the spatial inhomogeneity of the order
parameter do not exceed the ranges indicated above.
The inhomogeneous model systems we used are McMillan model systems
sandwiched between strong homeotropic anchoring walls;
we discuss these models and the effective fields in
\S~\ref{Discussion}. 

The phase diagrams of systems in nonzero external fields are shown in
Figs.\ref{fig:4phaseDiagrams} (b), (c) and (d) for various $h_{s}$ and
$h_{\sigma}$. 
These external fields change the phase diagram not only quantitatively
but also qualitatively.

We first discuss the effect of $h_{s}$.
If $h_{s}<h_{s}^{(\text{c})}=0.0104$, the application of $h_{s}$ does
not induce a topological change on the phase diagram, and increases both
$T_{IN}$ and $T_{NA}$ linearly.
We note that the triple point goes to high-$\alpha$ region by the
application of $h_{s}$.
Thus, under the field $h_{s}$, we have a parameter region in which the
nematic phase appears, even if the systems do not exhibit the nematic
phase under zero fields; i.e., the field induced non-spontaneous nematic
phase~\cite{Rosenblatt1981a,Lelidis1994} appears also in the McMillan
model. 
For $h_{s} > h_{s}^{\text{(c)}}$ and $h_{\sigma}=0$, the I-N coexisting
line entirely vanishes and thus only N-A coexisting line remains
(Fig.~\ref{fig:4phaseDiagrams}(b)).
The fact that I-N transition vanishes under a strong external field
$h_{s}$ was shown in ref.~\citen{wojtowicz} for $\alpha=0$, i.e., for
Maier-Saupe model.

The effect of $h_{\sigma}$ is observed on the small $\alpha$ region even
if $h_{\sigma}$ is very small.
The field $h_{\sigma}$ removes the second order transition part on the
N-A coexisting line (Fig.~\ref{fig:4phaseDiagrams}(c)). 
Thus the tricritical point turns to a critical point.
As $h_{\sigma}$ increases, N-A line moves to higher temperature region;
I-N line is not sensitive to $h_{\sigma}$.
If $h_{\sigma}$ is strong enough, even if $h_{s} >
h_{s}{}^{\text{(c)}}=0.0104$, a part of I-N line survives and it
terminates at the I-N critical point (Fig.~\ref{fig:4phaseDiagrams}(d)). 
As $h_{\sigma}$ increases, the temperature at I-N critical point
$T_{\text{IN}}{}^{\text{(c)}}$ increases, and $\alpha$ value at I-N
critical point $\alpha_{\text{IN}}{}^{\text{(c)}}$ decreases, so that
I-N line is elongated.
The application of $h_{s}$ also increases both of
$T_{\text{IN}}{}^{\text{(c)}}$ and $\alpha_{\text{IN}}{}^{\text{(c)}}$,
and thus shorten the I-N line.
The $\alpha_{\text{IN}}{}^{\text{(c)}}$ is sensitive to $h_{s}$, so that
the I-N line rapidly shrinks and vanishes for increasing $h_{s}$.
For $h_{s} > h_{s}{}^{\text{(c)}}$ and finite $h_{\sigma}$, there are
paths where phases may change from isotropic to nematic, and nematic to
smectic, without discontinuous transitions.
Furthermore, I-N line rapidly shrinks as $h_{s}$ increases, and N-A line
also rapidly shrinks as $h_{\sigma}$ increases.
Thus the bifurcated coexisting line becomes a single line which
terminates at a critical point.

\begin{figure}[t]
\begin{center}
\includegraphics[width=8.5cm]{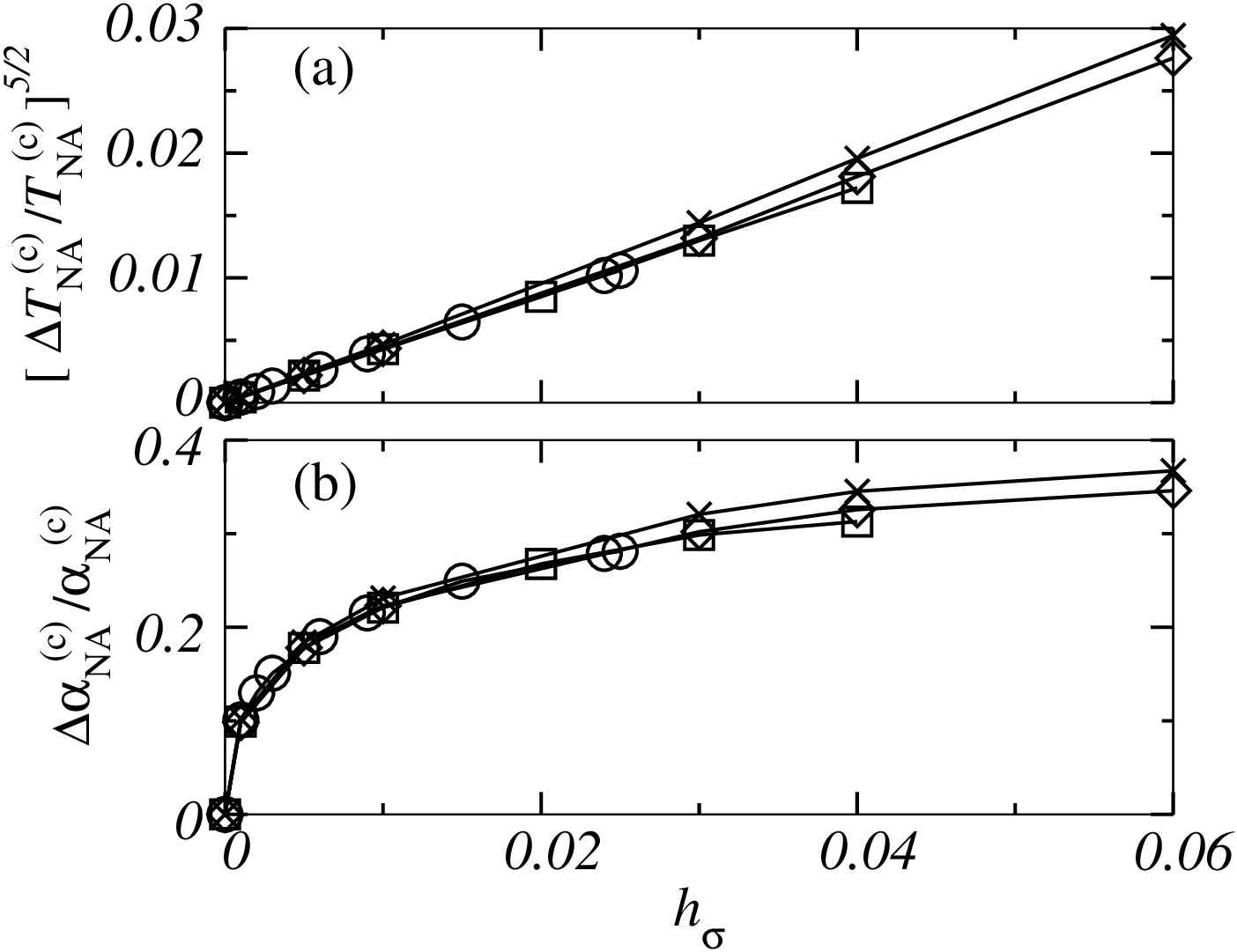}
\end{center}
\caption{
 The field $h_{\sigma}$ dependence of critical values:
 (a)critical temperature
 $\Delta T_{\text{NA}}{}^{\text{(c)}}=
 T_{\text{NA}}{}^{\text{(c)}}(h_{s}, h_{\sigma})
 -T_{\text{NA}}{}^{\text{(c)}}(h_{s}, 0)$,  
 and (b)critical parameter:
 $\Delta \alpha_{\text{NA}}{}^{\text{(c)}}=
 \alpha_{\text{NA}}{}^{\text{(c)}}(h_{s}, h_{\sigma})
 -\alpha_{\text{NA}}{}^{\text{(c)}}(h_{s}, 0)$.
 The symbols $\circ$, $\square$, $\diamond$, and $\times$ correspond to
 $h_{s}=0$, 0.011, 0.02, and 0.03, respectively.
 The curves for $h_{s}=0$ and 0.011 terminate at $h_{\sigma} \sim 0.025$
 and $\sim 0.04$, respectively;
 this is because N-A critical points vanish for large $h_{\sigma}$ since 
 $T_{\text{NA}}{}^{\text{(c)}}$ increases and the critical points merge
 with the I-N transition line.
 }
\label{fighsigmaVsTandAlpha}
\end{figure}

We estimate the effect of $h_{\sigma}$ on the N-A critical point
more quantitatively.
From our numerical calculations, the N-A critical temperature
$T_{\text{NA}}{}^{\text{(c)}}$ is obtained to depend on $h_{\sigma}$ as
\begin{equation}
 \frac{
 T_{\text{NA}}{}^{\text{(c)}}(h_{s}, h_{\sigma})
 -T_{\text{NA}}{}^{\text{(c)}}(h_{s}, 0)
 }
 {T_{\text{NA}}{}^{\text{(c)}}(h_{s}, 0)}
 =a h_{\sigma}^{2/5},
\end{equation}
as shown in Fig.\ref{fighsigmaVsTandAlpha}(a).
The $h_{s}$-dependence of the coefficient $a$ is weak;
the coefficients $a$ are 0.423, 0.431, 0.458 and 0.491 for 
$h_{s}=0, 0.011, 0.020$ and 0.030, respectively. 
The $h_{s}$ dependence of N-A critical point is approximately linear,
i.e., for $h_{\sigma}=0$, 
\begin{equation}
 \frac{
 T_{\text{NA}}{}^{\text{(c)}}(h_{s}, 0)
 -T_{\text{NA}}{}^{\text{(c)}}(0, 0)}
 {T_{\text{NA}}{}^{\text{(c)}}(0, 0)}
 =b h_{s},
\end{equation}
where $b=1.88$.
Thus $h_{s}$ and $h_{\sigma}$ dependence of
$T_{\text{NA}}{}^{\text{(c)}}$ is completed as 
\begin{equation}
 \frac{
 T_{\text{NA}}{}^{\text{(c)}}(h_{s}, h_{\sigma})
 -T_{\text{NA}}{}^{\text{(c)}}(0, 0)}
 {T_{\text{NA}}{}^{\text{(c)}}(0, 0)}
 =
 b h_{s}
 +a (1+b h_{s}) h_{\sigma}^{2/5}.
\end{equation}
If $h_{s}$ increases, the $T_{\text{NA}}{}^{\text{(c)}}$ increases up to
I-N transition temperature, then N-A line vanishes.
The $h_{\sigma}$-dependence of N-A critical value of $\alpha$,
$\alpha_{\text{NA}}{}^{\text{(c)}}$, seems to be scalable for
several $h_{s}$ as shown in Fig.\ref{fighsigmaVsTandAlpha}(b).
However, in the present stage, it is difficult to obtain a simple
function form of scaled $\alpha_{\text{NA}}{}^{\text{(c)}}$.

\begin{figure}
 \begin{center}
  \includegraphics[width=8.5cm]{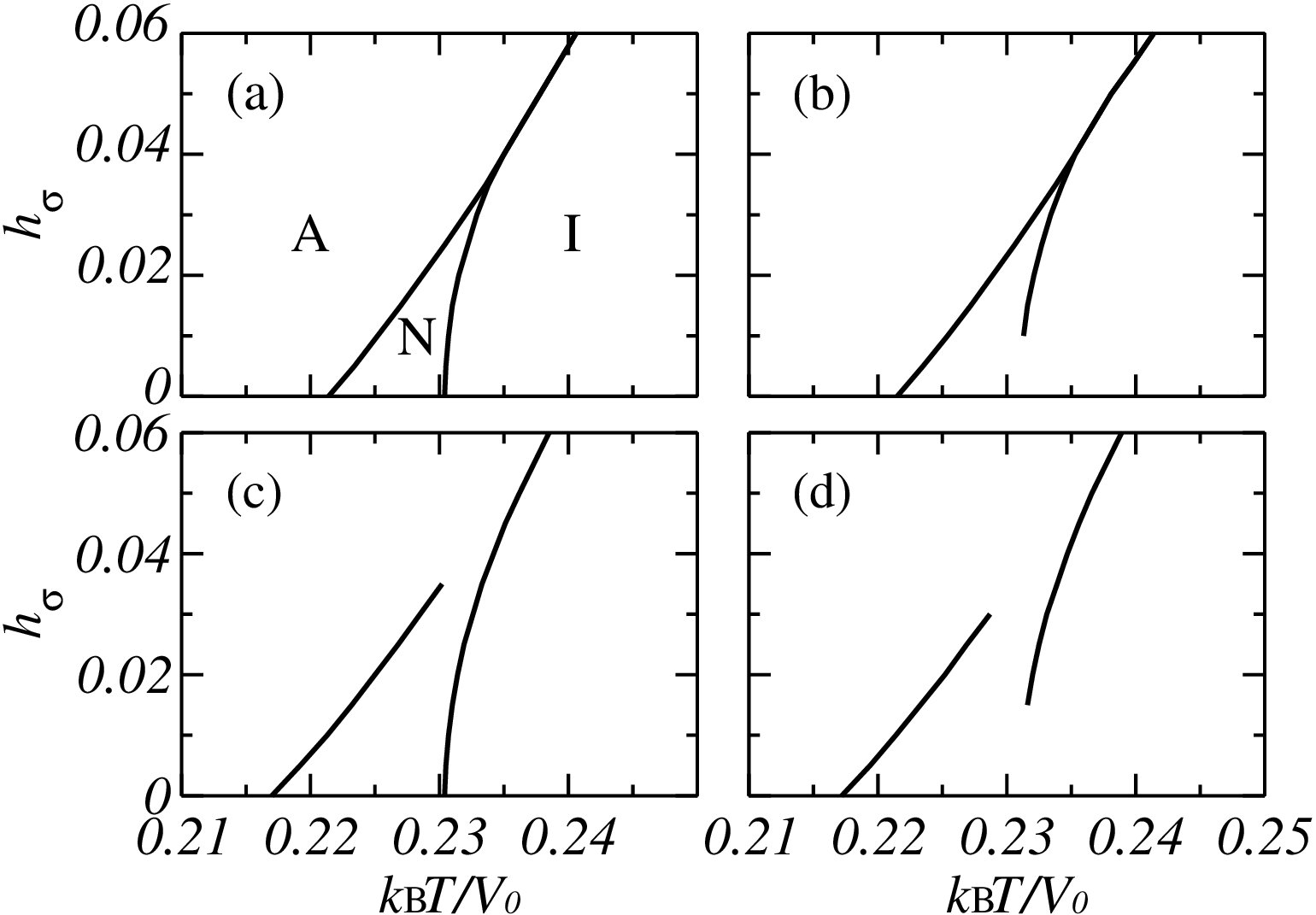}
 \end{center}
 \caption{
 The temperature-field ($T$-$h_{\sigma}$) phase diagrams for
 (a)$\alpha=0.975, h_{s}=0.01<h_{s}^{\text(c)}$,
 (b)$\alpha=0.975, h_{s}=0.0106>h_{s}^{\text(c)}$,
 (c)$\alpha=0.93, h_{s}=0.01$, and
 (d)$\alpha=0.93, h_{s}=0.0106$.
 The labels I, N and A in (a) denote the isotropic, nematic and smectic
 A phases, respectively.
 }
 \label{figHsigmaVsT}
\end{figure}

In order to show clearly the effect of $h_{\sigma}$ on the phase
behaviour, we show $h_{\sigma}$-$T$ phase diagrams in
Fig.~\ref{figHsigmaVsT}. 
In the system with $\alpha=0.975$, if the external field $h_{s}$ is
weak, three phases are divided by first order transition lines
(Fig.~\ref{figHsigmaVsT}(a)). 
The transition temperature $T_{\text{NA}}$ increases almost linearly to
$h_{\sigma}$. 
On the other hand, $h_{\sigma}$-dependence of $T_{\text{IN}}$ is weak.
In fact, in the vicinity of $h_{\sigma}=0$, $T_{\text{IN}}$ does not
depend on $h_{\sigma}$.
In analogy to the Clapeyron equation, we obtain
\begin{equation}
 \frac{d T_{\text{IN}}}{d h_{\sigma}}
  =-\frac{\Delta(\partial F/\partial h_{\sigma})}{\Delta F_{T}}
  =\frac{\Delta \sigma}{\Delta S}, 
\end{equation}
where $\Delta \sigma$ and $\Delta S$ are the changes of, respectively,
$\sigma$ and entropy per molecule at the I-N transition.
Thus, $d T_{\text{IN}}/d h_{\sigma}$ is zero since $\sigma$ is zero in
both isotropic and nematic phases at $h_{\sigma}=0$. 
By the application of $h_{\sigma}$, the smectic order is induced in
isotropic and nematic phases, so that the change in $\sigma$ is finite
and $T_{\text{IN}}$ becomes dependent on $h_{\sigma}$. 
If the field $h_{s}$ exceeds $h_{s}{}^{\text{(c)}}$, in the system with
$\alpha=0.975$, the I-N transition turns to be continuous at
$h_{\sigma} \sim 0$.
But the application of $h_{\sigma}$ recovers the I-N line thus the system
again undergoes I-N transition (Fig.~\ref{figHsigmaVsT}(b)).
For $h_s<h_{s}{}^{\text{(c)}}$ and small $\alpha$, \textit{e.g.}
$\alpha=0.93$, the N-A transition turns to be continuous under high
$h_{\sigma}$; 
thus N-A critical point appears (Fig.~\ref{figHsigmaVsT}(c)).
If $h_{s}$ exceeds $h_{s}{}^{\text{(c)}}$ in $\alpha=0.93$ system, I-N
transition at low $h_{\sigma}$ becomes a continuous change without
transition. 
Then the system has paths from isotropic phase to smectic phase without
discontinuous transition (Fig.~\ref{figHsigmaVsT}(d)).

\section{Summary and Discussions}\label{Discussion}
We have studied the phase diagram of the McMillan model system under
external fields $h_{s}$, and $h_{\sigma}$.
By the application of $h_{\sigma}$, second order transition line on N-A
line at low $\alpha$ region vanishes and N-A transition turns to be
smooth.
The I-N line vanishes under an external field $h_{s}$ higher than
$h_{s}{}^{\text{(c)}}$, however $h_{\sigma}$ recovers the nematic order
and I-N coexisting line.

The fictitious field $h_{\sigma}$, which we mainly discussed in the
present paper, is resulted from the spatial inhomogeneity of the order
parameter $\sigma$. 
As an concrete example, let us consider a thin system sandwiched by two
parallel walls and assume that the walls force liquid crystalline
molecules at wall surfaces to stick perpendicular to the
walls(homeotropic anchoring). 
In the following, we discretise the system into $N$ layers parallel to
the walls, and assume the $n$th layer has order parameters $s^{(n)}$ and
$\sigma^{(n)}$. 
In the mean field theory, the inhomogeneity due to walls changes the
mean field contributions in the arguments of
eqs.\eqref{eq:ssEqWithField}; 
\textit{e.g.}, $V_{0}s$ turns to
$V_{0}'s^{(n)} + V_{0}''s^{(n-1)} + V_{0}''s^{(n+1)}$ with constants
$V_{0}'$ and $V_{0}''$ denoting intralayer and interlayer interactions,
respectively.
Then we may write the self-consistent equation as
\begin{subequations}
 \begin{align}
 s^{(n)}= 
 I(\beta,
 \beta h_{s}^{(n)}+\beta V_{0} s^{(n)},
 \beta h_{\sigma}^{(n)}+\beta V_{0} \alpha \sigma^{(n)}),
 \label{eq:scEq2-1}\\
 \sigma^{(n)}=
 J(\beta,
 \beta h_{s}^{(n)}+\beta V_{0} s^{(n)},
 \beta h_{\sigma}^{(n)}+\beta V_{0} \alpha \sigma^{(n)}),
 \label{eq:scEq2-2}
 \end{align}\label{eq:scEq2}
\end{subequations}
where the $h_{s}^{(n)}$ and $h_{\sigma}^{(n)}$ are defined by
\begin{subequations}
 \begin{align}
  h_{s}^{(n)}=V_{0}''(s^{(n-1)}-2s^{(n)}+s^{(n+1)}),
  \label{eq:effectiveFields1}\\
  h_{\sigma}^{(n)}=V_{0}''(\sigma^{(n-1)}-2\sigma^{(n)}+\sigma^{(n+1)}),
  \label{eq:effectiveFields2}
 \end{align}\label{eq:effectiveFields}
\end{subequations}
and the boundary conditions are $s^{(1)}=s^{(N)}=1$ and
$\sigma^{(1)}=\sigma^{(N)}=1$.
In eqs.\eqref{eq:scEq2} and \eqref{eq:effectiveFields} we redefined
$V_{0}=V_{0}'+2V_{0}''$ in order to emphasise the equivalence of
eqs.\eqref{eq:ssEqWithField} and \eqref{eq:scEq2} in the homogeneous
limit. 
We summarised the effect of inhomogeneity of the order parameters into
$h_s^{(n)}$ and $h_{\sigma}^{(n)}$.
From the similarity between eqs.~\eqref{eq:ssEqWithField} and
\eqref{eq:scEq2}, we may call the variables $h_{s}^{(n)}$ and
$h_{\sigma}^{(n)}$ the effective fields corresponding to $h_{s}$ and
$h_{\sigma}$, respectively.
The stability condition for the inhomogeneous system is minimizing the
work done by the symmetry breaking fields on the whole system, not the
work on the each layer.
Except for the difference in the stability conditions, each layer of the
inhomogeneous thin system under effective fields may be viewed as a
homogeneous bulk system under external fields which depend on position
and on temperature.
From this point of view, we may analyse the thermal behaviour of the
thin system by observing the locus of the fields on the bulk phase
diagram of the field versus temperature plane.
In fact, such an analysis is shown to be useful in order to uncover the
mechanism of smectic A-C$^{*}$ phase transition~\cite{yamashita2003a},
and I-N phase transition of Maier-Saupe
model~\cite{yamashita2003b,Yasen2004}.
For thin system version of the model we analysed in the present paper,
we have investigated the phase transitions and thermal
behaviour of physical quantities by utilising the phase diagram
obtained in the present paper, which will be reported in a near
future~\cite{torikai}. 

This study was partly supported by a Grant-in-Aid from the Ministry of 
Education, Culture, Sports, Science and Technology (No. 14540355).


\begin{thebibliography}{99}
 \bibitem{wojtowicz} P. J. Wojtowicz and P. Sheng:
	 Phys. Lett. \textbf{48A} (1974) 235.
 \bibitem{Lelidis1993b} I. Lelidis, and G. Durand:
	 Phys. Rev. E \textbf{48} (1993) 3822.
 \bibitem{Rosenblatt1981a} C. Rosenblatt:
	 Phys. Lett. A \textbf{83} (1981) 221.
 \bibitem{Lelidis1994} I. Lelidis, and G. Durand:
	 Phys. Rev. Lett. \textbf{73} (1994) 672.
 \bibitem{Rosenblatt1981b} C. Rosenblatt:
	 Journal de Physique Letters \textbf{42} (1981) 9.
 \bibitem{Hama} H. Hama:
	 J. Phys. Soc. Jpn. \textbf{54} (1985) 2204.
 \bibitem{deGennes} P. G. de Gennes:
	 \textit{The Physics of Liquid Crystals} (Clarendon, Oxford,
	 1973) 
 \bibitem{Lelidis1996} I. Lelidis and G. Durand:
	 Journal de Physique II \textbf{6} (1996) 1359.
 \bibitem{yamashita2003a} M. Yamashita:
	 J. Phys. Soc. Jpn. \textbf{72} (2003) 2421.
 \bibitem{McMillan} W. L. McMillan:
	 Phys. Rev. A \textbf{4} (1971) 1238.
 \bibitem{Callen} H. B. Callen:
	 \textit{Thermodynamics and an Introduction to Thermostatistics}
	 (John Wiley \& Sons, New York, 1985) 2nd ed.
 \bibitem{Nakano} H. Nakano and M. Hattori:
	 Prog. Theor. Phys. \textbf{49} (1973) 1752.
 \bibitem{yamashita2003b} M. Yamashita:
	 J. Phys. Soc. Jpn. \textbf{72} (2003) 1682.
 \bibitem{Yasen2004} M. Yasen, M. Torikai and M. Yamashita:
	 submitted to J. Phys. Soc. Jpn.
 \bibitem{torikai} M. Torikai and M. Yamashita:
	 in preparation for publication.
\end{thebibliography}
\end{document}